\documentclass[a4paper,11pt, onesided]{article}

\usepackage[ansinew]{inputenc}

\usepackage{amsfonts}
\usepackage{amsmath}
\usepackage{amsthm}

\usepackage{mathrsfs}
\usepackage{amssymb}

\newtheorem{theoremp4}{Theorem}
\newtheorem{lemmap4}[theoremp4]{Lemma}
\newtheorem{corollaryp4}[theoremp4]{Corollary}

\begin{document}

\title{Conserved Matter Superenergy Currents for Orthogonally Transitive Abelian $G_2$ Isometry Groups}
    \author{Ingemar Eriksson\\
        Matematiska institutionen, Linköpings universitet, \\
        SE-581 83 Linköping, Sweden \\
        ineri@mai.liu.se}
    \date{June 27, 2007}

    \maketitle

\begin{abstract}
    In a previous paper we showed that the electromagnetic superenergy tensor, the Chevreton tensor, gives rise to a
    conserved current when there is a hypersurface orthogonal Killing vector present. In addition, the current is
    proportional to the Killing vector. The aim of this paper is to extend this result to the case when we have a
    two-parameter Abelian isometry group that acts orthogonally transitive on non-null surfaces.
    It is shown that for four-dimensional Einstein-Maxwell theory with a source-free electromagnetic field, the
    corresponding superenergy currents lie in the orbits of the group and are conserved. A similar result is also shown
    to hold for the trace of the Chevreton tensor and for the Bach tensor, and also in Einstein-Klein-Gordon theory for
    the superenergy of the scalar field. This links up well with the fact that the Bel tensor has these properties and
    the possibility of constructing conserved mixed currents between the gravitational field and the matter fields.
\end{abstract}


\section{Introduction}

In this paper we continue the investigation of conservation laws for the electromagnetic superenergy tensor, the
Chevreton tensor, in Einstein-Maxwell spacetimes. In a previous paper \cite{Eriksson2006} we showed that this tensor
gives rise to a conserved current whenever there is a hypersurface orthogonal Killing vector present, i.e., if the
Killing vector $\xi_a$ satisfies $\xi_{[a}\nabla_b\xi_{c]} = 0$, then
\begin{align}
    H_{abcd} \xi^b\xi^c\xi^d &= \omega\xi_a,    &       \nabla^a(H_{abcd}\xi^b\xi^c\xi^d) &= 0,
\end{align}
where $H_{abcd}$ is the Chevreton tensor \cite{Chevreton1964, Senovilla2000},
\begin{align}
     H_{abcd} = & -\frac{1}{2}( \nabla_a F_{ce} \nabla_b F_d{}^e + \nabla_b F_{ce} \nabla_a F_d{}^e +
                                \nabla_c F_{ae} \nabla_d F_b{}^e + \nabla_d F_{ae} \nabla_c F_b{}^e ) \nonumber\\
                & +\frac{1}{2}( g_{ab}\nabla_f F_{ce} \nabla^f F_d{}^e + g_{cd}\nabla_f F_{ae} \nabla^f F_b{}^e )
                        +\frac{1}{4}( g_{ab}\nabla_c F_{ef} \nabla_d F^{ef} + g_{cd}\nabla_a F_{ef} \nabla_b F^{ef} )\nonumber\\
                & -\frac{1}{4} g_{ab}g_{cd}\nabla_e F_{fg} \nabla^e F^{fg}.
\end{align}
This result holds in four-dimensional Einstein-Maxwell spacetimes with a source-free electromagnetic field, $F_{ab}$,
that inherits the symmetry of the spacetime. A similar situation to when there is a hypersurface orthogonal Killing
vector present occurs when there exists a two-parameter isometry group whose surfaces of transitivity are (locally)
orthogonal to a family of $2$-surfaces. According to Frobenius's theorem the two Killing vectors generating the group,
$\xi_a$ and $\eta_a$, then satisfies $\xi_{[a} \eta_b \nabla_c \xi_{d]} = 0 = \xi_{[a} \eta_b \nabla_c \eta_{d]}$. We
show in this paper, theorem \ref{BigTheorem}, that when the two Killing vectors commute (i.e., the isometry group is
Abelian) and form surfaces that are non-null, then the Chevreton tensor again gives rise to conserved currents,
\begin{align}
    H_{abcd} \xi^{Ib}\xi^{Jc} \xi^{Kd} &= \omega_{IJK} \xi_a + \Omega_{IJK} \eta_a   &   \nabla^a(H_{abcd} \xi^{Ib}\xi^{Jc} \xi^{Kd}) &= 0,
\end{align}
where $I,J,K = 1,2$ and $\xi^1_a = \xi_a$ and $\xi^2_a = \eta_a$. We also show that similar results hold for the trace
of the Chevreton tensor and for the Bach tensor.

This result is interesting not only because it gives conserved quantities for the electromagnetic field, but also
because it gives further support to the possibility of creating conserved currents between the electromagnetic field
and the gravitational field at the superenergy level.

The Bel-Robinson tensor \cite{Bel1958, Bel1959},
\begin{align} \label{BRTensor}
  T_{abcd} =& C_{aecf}C_b{}^e{}_d{}^f + C_{aedf}C_b{}^e{}_c{}^f - \frac{1}{2}g_{ab}C_{efcg}C^{ef}{}_d{}^g\nonumber\\
            & -\frac{1}{2}g_{cd}C_{aefg}C_b{}^{efg} + \frac{1}{8}g_{ab}g_{cd}C_{efgh}C^{efgh},
\end{align}
is a good candidate for representing gravitational energy since it satisfies the Dominant Property \cite{Bergqvist1999, Senovilla2000} and is
divergence-free in vacuum. When matter is present, however, neither the Bel-Robinson tensor, nor the Bel tensor,
\begin{align} \label{BelTensor}
  B_{abcd} =& R_{aecf}R_b{}^e{}_d{}^f + R_{aedf}R_b{}^e{}_c{}^f - \frac{1}{2}g_{ab}R_{efcg}R^{ef}{}_d{}^g\nonumber\\
            & -\frac{1}{2}g_{cd}R_{aefg}R_b{}^{efg} + \frac{1}{8}g_{ab}g_{cd}R_{efgh}R^{efgh},
\end{align}
are divergence-free in general. However, there are some cases when it is still possible to construct conserved currents
for the gravitational field at the superenergy level. Lazkoz, Senovilla, and Vera \cite{Lazkoz2003} have shown that the
Bel tensor gives rise to independently conserved currents for general spacetimes when there is a hypersurface
orthogonal Killing vector present or when there are two commuting Killing vectors that act orthogonally transitive on
non-null surfaces present. In the first case we have the current
\begin{align}
    B_{abcd} \xi^b\xi^c \xi^d &= \omega \xi_a,    &   \nabla^a(B_{abcd}\xi^b\xi^c\xi^d) &= 0,
\end{align}
and in the second case the four currents
\begin{align}
    B_{a(bcd)} \xi^{Ib}\xi^{Jc} \xi^{Kd} &= \omega_{IJK} \xi_a + \Omega_{IJK} \eta_a,   &   \nabla^a(B_{a(bcd)} \xi^{Ib}\xi^{Jc} \xi^{Kd}) &= 0.
\end{align}
Also, Senovilla \cite{Senovilla2000} has shown that for Einstein-Klein-Gordon theory, it is possible to construct a
mixed conserved superenergy current between the gravitational field and the scalar field when there is a Killing vector
present,
\begin{align}
    \nabla^a\left( (B_{abcd} + S_{abcd}) \xi^b\xi^c\xi^d \right) = 0,
\end{align}
where $S_{abcd}$ is the superenergy of the scalar field. When the above isometries are present this breaks up into two
separate conserved currents and we show for completeness here and in \cite{Eriksson2006} that the currents constructed
from the superenergy of the scalar field also lie in the orbits of the isometry groups.

We hope that it would be possible to construct a similar conserved current between the gravitational field and the
electromagnetic field. Senovilla \cite{Senovilla2000} has shown that this is possible in the case of propagation of
discontinuities of the fields. For the general case it is not known, but the results of this paper further support
 that this might be the case. Also, in the spacetime we use as an example of our results, we do have mixed conserved
currents.

In the proofs we have opted for expanding the tensors in a basis where the two Killing vectors are taken as two of
the basis vectors. It is also possible to take exterior products with the surface element
$\xi_{[a}\eta_{b]}$ and using expressions like
$2 \xi_{[a}\eta_b\nabla_{c]}\xi_d = -\eta_d \xi_{[a}\nabla_b\xi_{c]} + \xi_d \eta_{[a}\nabla_b\xi_{c]}$, but this
approach seems to require quite a lot of extra effort.


\section{Conventions and some results}
We assume that our spacetime is a four-dimensional manifold equipped with a metric of signature $-2$. We define
the Riemann tensor by
\begin{align}
    (\nabla_a\nabla_b - \nabla_b\nabla_a)v_c = -R_{abcd}v^d.
\end{align}
The Einstein equations are
\begin{align} \label{EinsteinEqs}
    R_{ab} - \frac{1}{2}Rg_{ab} + \Lambda g_{ab} = -T_{ab}.
\end{align}
We will keep the cosmological constant $\Lambda$ throughout the calculations. If $\xi_a$ is a Killing vector, then
$\nabla_a\xi_b = -\nabla_b\xi_a$ and \cite{Wald1984},
\begin{align}\label{Killing-riemann}
    \nabla_a\nabla_b\xi_c = R_{bcad}\xi^d.
\end{align}
We also note that the Lie derivative along a Killing vector commutes with the covariant derivative \cite{Yano1955},
\begin{align}\label{LieCovariant}
    [\pounds_\xi,\nabla_a] T^{b_1\ldots b_i}{}_{c_1\ldots c_j}= 0.
\end{align}
We assume that we have two commuting Killing vectors $\xi^a$ and $\eta^a$ that act orthogonally transitive on non-null
surfaces \cite{Stephani2003},
\begin{align}
    [\xi,\eta] = \xi^b\nabla_b \eta_a - \eta^b\nabla_b\xi_a &= 0, \\
    \xi_{[a}\eta_b \nabla_c \xi_{d]} = \xi_{[a}\eta_b \nabla_c \eta_{d]} &= 0,\\
    \xi_{[a}\eta_{b]}\xi^a\eta^b &\neq 0.
\end{align}
If we take a basis consisting of $\xi_a$, $\eta_a$, $s_a$, and $t_a$, where $s_a$ and $t_a$ are orthogonal to $\xi_a$
and $\eta_a$, we can write
\begin{align}
    \nabla_a\xi_b = C_1\xi_{[a}\eta_{b]} + \xi_{[a}(C_2 s_{b]}+C_3t_{b]})+\eta_{[a}(C_4 s_{b]}+C_5t_{b]}) + C_6 s_{[a}t_{b]}.
\end{align}
Taking an exterior product with $\xi_{[c}\eta_{d]}$ gives $C_6=0$ and by contracting with $\xi^a\eta^b$ we get
\begin{align} \label{KillingKillingNablaKilling}
    \xi^a\eta^b\nabla_a\xi_b = -\xi^a\eta^b\nabla_b\xi_a = -\xi^a\xi^b\nabla_b\eta_a = 0 = C_1 \xi_{[a}\eta_{b]}\xi^a\eta^b,
\end{align}
so $C_1 = 0$. The structure is the same for $\nabla_a\eta_b$ and we can write
\begin{align}
    \nabla_a\xi_b  &= 2\xi_{[a}x_{b]} + 2\eta_{[a}y_{b]}, \\
    \nabla_a\eta_b &= 2\xi_{[a}v_{b]} + 2\eta_{[a}w_{b]},
\end{align}
where $x_a, y_a, v_a$, and $w_a$ are orthogonal to $\xi_a$ and $\eta_a$. We will often write this as
\begin{align} \label{Nablakilling-expanded}
    \nabla_a\xi^I_b  &= 2\xi_{[a}x^I_{b]} + 2\eta_{[a}y^I_{b]},
\end{align}
where $I = 1,2$ and $x^1_a = x_a$, $x^2_a = v_a$, $y^1_a = y_a$, and $y^2_a = w_a$. Via Einstein's equations
the energy-momentum tensor satisfies \cite{Carter1969, Lazkoz2003}
\begin{align}
    \xi_{[a}\eta_b T_{c]d}\xi^d = \xi_{[a}\eta_b T_{c]d}\eta^d &= 0,
\end{align}
which implies
\begin{align}\label{Killing-energymomentum}
    \xi^{Ib} T_{ab}  &= \alpha^I \xi_a + \beta^I \eta_a,
\end{align}
where $\alpha^1 = \alpha$, $\alpha^2 = \gamma$, $\beta^1 = \beta$, and $\beta^2 = \delta$. By taking the
Lie derivatives of this equation with respect to $\xi_a$ and $\eta_a$ we have that
\begin{align} \label{Killing-scalars}
    \xi^a \nabla_a \alpha^I &= 0,  &        \eta^a \nabla_a \alpha^I &= 0, \nonumber\\
    \xi^a \nabla_a \beta^I &= 0,  &        \eta^a \nabla_a \beta^I &= 0.
\end{align}
When there is more than one matter field present this, in general, only applies to the total energy-momentum tensor.
Here, as well as throughout the text, proportionality factors like $\alpha$ in $\alpha\xi_a$ are generally non-constant
scalar functions.


\section{Einstein-Klein-Gordon theory}
In this section we show that in Einstein-Klein-Gordon theory the superenergy tensor of the scalar (Klein-Gordon) field
gives rise conserved currents for Killing vectors that generate an Abelian two-parameter group of isometries
that act orthogonally transitive on non-null surfaces. It has previously been shown that the Bel tensor in combination
with the superenergy tensor of the scalar field gives rise to conserved currents for Killing vectors \cite{Senovilla2000} and
that for this kind of symmetry, or for hypersurface orthogonal Killing vectors, the Bel tensor gives rise to
independently conserved currents that lie in the orbits of the group \cite{Lazkoz2003}. Hence, the superenergy currents
for the scalar field are also independently conserved, and we show here and in \cite{Eriksson2006} for completeness
that these currents also lie in the orbits of the group.

The energy-momentum tensor in Einstein-Klein-Gordon theory is given by
\begin{align} \label{KGem}
  T_{ab} = -\nabla_a\phi\nabla_b\phi +
  \frac{1}{2}g_{ab}( \nabla_c\phi\nabla^c\phi + m^2\phi^2),
\end{align}
where the scalar field, $\phi$, satisfies the Klein-Gordon equation,
$\nabla^c\nabla_c\phi = m^2\phi$. The superenergy tensor of the scalar field
is given by \cite{Senovilla2000}
\begin{align}
  S_{abcd} =& \nabla_a\nabla_c\phi\nabla_b\nabla_d\phi + \nabla_a\nabla_d\phi\nabla_b\nabla_c\phi
                - g_{ab}( \nabla_c\nabla^e\phi\nabla_d\nabla_e\phi + m^2 \nabla_c\phi\nabla_d\phi )\nonumber\\
            &-g_{cd}( \nabla_a\nabla^e\phi\nabla_b\nabla_e\phi + m^2 \nabla_a\phi\nabla_b\phi ) \nonumber\\
            &    +\frac{1}{2}g_{ab}g_{cd}( \nabla_e\nabla_f\phi\nabla^e\nabla^f\phi + 2m^2\nabla_e\phi\nabla^e\phi + m^4\phi^2).
\end{align}
It has the following symmetries, $S_{abcd} = S_{(ab)(cd)} = S_{cdab}$. We can construct the following currents,
\begin{align} \label{KGtemp}
    S_{abcd}\xi^{Ib}\xi^{Jc}\xi^{Kd} =& \left(
            \nabla_a\nabla_c\phi\nabla_b\nabla_d\phi + \nabla_a\nabla_d\phi\nabla_b\nabla_c\phi \right)\xi^{Ib}\xi^{Jc}\xi^{Kd}\nonumber\\
            &- \frac{1}{3} \xi^{Jc}\xi^K_c \left( \nabla_a\nabla^e\phi\nabla_b\nabla_e\phi + m^2\nabla_a\phi\nabla_b\phi\right)\xi^{Ib}
                +\omega\xi_a + \Omega\eta_a,
\end{align}
where the scalar functions $\omega$ and $\Omega$ are used to collect proportionality factors of $\xi_a$ and $\eta_a$.
If the scalar field is massive, $m\neq 0$, it will have a vanishing Lie derivative \cite{Senovilla2000},
$\pounds_{\xi^I}\phi = \xi^{Ia}\nabla_a\phi=0$, so by using Leibniz rule and expanding with
(\ref{Nablakilling-expanded}) we have that
\begin{align}
  &  \left( \nabla_a\nabla^e\phi\nabla_b\nabla_e\phi + m^2\nabla_a\phi\nabla_b\phi\right)\xi^{Ib} =
            \nabla_a\nabla^e\phi\nabla_b\nabla_e\phi\xi^{Ib} = \nabla_a\nabla^e\phi\nabla_e\nabla_b\phi\xi^{Ib} \nonumber\\
 =& -\nabla_a\nabla^e\phi \nabla_b\phi \nabla_e\xi^{Ib} =
           -\xi^e\nabla_a\nabla_e\phi x^{Ib}\nabla_b\phi - \eta^e\nabla_a\nabla_e\phi y^{Ib}\nabla_b\phi \nonumber\\
 =& \nabla_a\xi^e\nabla_e\phi x^{Ib}\nabla_b\phi + \nabla_a\eta^e\nabla_e\phi y^{Ib}\nabla_b\phi
    = \omega\xi_a + \Omega\eta_a.
\end{align}
For the other type of term present in (\ref{KGtemp}) we similarly have that
\begin{align}
        \nabla_a\nabla_c\phi\nabla_b\nabla_d\phi \xi^{Ib}\xi^{Jc}\xi^{Kd} = -\nabla_a\xi^{Jc}\nabla_c\phi\nabla_b\nabla_d\phi \xi^{Ib}\xi^{Kd}
        = \omega\xi_a + \Omega\eta_a.
\end{align}
Hence, for a massive scalar field the superenergy currents will lie in the orbits of the group,
$S_{abcd}\xi^{Ib}\xi^{Jc}\xi^{Kd} = \omega\xi_a + \Omega\eta_a$.

In the massless case, $m=0$, the scalar field satisfies $\pounds_{\xi^I}\phi = \xi^{Ia}\nabla_a\phi = C_{\xi^I}$, where
$C_{\xi^I}$ is a constant. Here we also have from (\ref{Killing-energymomentum}) and (\ref{KGem}) that
$\nabla_a\phi = \alpha'\xi_a+\beta'\eta_a$. The calculations are similar to the massive case and the conclusion the
same.

To prove that the currents are divergence-free, we note that the Lie derivative commutes with covariant derivatives for
Killing vectors, so the superenergy tensor has vanishing Lie derivative. Since the Killing vectors commute, the
Lie derivative of the currents therefore vanishes. We have proven

\begin{theoremp4}
    For Einstein-Klein-Gordon spacetimes, possibly with a cosmological constant $\Lambda$, which admit an Abelian
    two-parameter isometry group that act orthogonally transitive on non-null surfaces, the superenergy tensor of the
    scalar field gives rise to conserved currents that lie in the orbits of the group,
    \begin{align}
        S_{abcd}\xi^{Ib}\xi^{Jc}\xi^{Kd} &= \omega_{IJK}\xi_a + \Omega_{IJK}\eta_a,   &
        \nabla^a\left(S_{abcd}\xi^{Ib}\xi^{Jc}\xi^{Kd}\right) &= 0,
    \end{align}
    where, in general, the proportionality factors $\omega_{IJK}$ and $\Omega_{IJK}$ will be non-constant.
\end{theoremp4}

Note, we have here assumed a four-dimensional spacetime, but the expansion of (\ref{Nablakilling-expanded}) is similar
in the $n$-dimensional case and this result thus holds in $n$ dimensions as well.


\section{Einstein-Maxwell theory}
In this section we will show that if a four-dimensional Einstein-Maxwell spacetime, possibly with a cosmological
constant $\Lambda$, admits an Abelian two-parameter group of isometries that act orthogonally transitive on non-null
surfaces, then the corresponding Chevreton currents constructed from the Killing vectors of the group will lie in the
orbits of the group and will be conserved. It is also shown that this also holds for the trace of the Chevreton tensor
and for the Bach tensor. The electromagnetic field is assumed to be source-free. For a null electromagnetic field we
will assume that it inherits the symmetries of the spacetime.

The electromagnetic field is described by the Maxwell tensor, $F_{ab} = -F_{ba}$, which in source-free regions
satisfies
\begin{align}
    \nabla^a F_{ab} = 0, \qquad \nabla_{[a}F_{bc]} = 0.
\end{align}
The energy momentum tensor is given by
\begin{align} \label{EM-energymomentum}
    T_{ab} = -F_{ac}F_b{}^c + \frac{1}{4}g_{ab}F_{cd}F^{cd}.
\end{align}
The Ricci scalar, $R$, satisfies $R = 4\Lambda$, where $\Lambda$ is the cosmological constant. From
(\ref{Killing-energymomentum}) we have that
\begin{align} \label{Killing-energymomentum-prime}
    \xi^{Ib} F_{ac}F_b{}^c = \alpha'^I \xi_a + \beta'^I \eta_a.
\end{align}
Generally, the Lie derivative of the electromagnetic field in four-dimensional Einstein-Maxwell theory satisfies for
any Killing vector $\xi^a$ \cite{Michalski1975, Wainwright1976}
\begin{align}
    \pounds_\xi F_{ab} = \xi^c\nabla_c F_{ab} + F_{cb}\nabla_a\xi^c + F_{ac}\nabla_b\xi^c = \Psi \stackrel{*}F_{ab},
\end{align}
where $\stackrel{*}F_{ab}$ is the Hodge dual of $F_{ab}$ and $\Psi$ is a constant for non-null fields and satisfies
$l_{[a} \nabla_{b]}\Psi = 0$ for null fields, where $l_a$ is the repeated principal null direction of the
field.\footnote{It was erroneously stated in \cite{Eriksson2006} that $\Psi$ is always constant, though this did not
interfere with the calculations since the electromagnetic field was assumed to inherit the symmetry of the spacetime.}
If $\Psi$ is zero, then the electromagnetic field is said to inherit the symmetry of the spacetime. It has been shown
that for an
Abelian two-parameter group of isometries that acts orthogonally transitive on non-null surfaces, a non-null
electromagnetic field inherits those symmetries of the spacetime \cite{Michalski1975}. In the case of a null
electromagnetic field we will assume that it inherits the symmetries. Hence,
$\pounds_{\xi^I} F_{ab} = 0$, or
\begin{align} \label{EM-Liederivative}
    \xi^{Ic}\nabla_c F_{ab} = - F_{cb}\nabla_a\xi^{Ic} - F_{ac}\nabla_b\xi^{Ic}.
\end{align}
The basic superenergy tensor of the electromagnetic field is given by \cite{Senovilla2000}
\begin{align}
     E_{abcd} =& - \nabla_a F_{ce} \nabla_b F_d{}^e - \nabla_b F_{ce} \nabla_a F_d{}^e + g_{ab}\nabla_f F_{ce} \nabla^f F_d{}^e \nonumber\\
               &+ \frac{1}{2} g_{cd}\nabla_a F_{ef} \nabla_b F^{ef} - \frac{1}{4} g_{ab}g_{cd}\nabla_e F_{fg} \nabla^e F^{fg}.
\end{align}
The Chevreton tensor is defined as $H_{abcd} = \frac{1}{2}(E_{abcd}+E_{cdab})$, or
\begin{align} \label{Chevreton}
     H_{abcd} = & -\frac{1}{2}( \nabla_a F_{ce} \nabla_b F_d{}^e + \nabla_b F_{ce} \nabla_a F_d{}^e +
                                \nabla_c F_{ae} \nabla_d F_b{}^e + \nabla_d F_{ae} \nabla_c F_b{}^e ) \nonumber\\
                & +\frac{1}{2}( g_{ab}\nabla_f F_{ce} \nabla^f F_d{}^e + g_{cd}\nabla_f F_{ae} \nabla^f F_b{}^e )
                        +\frac{1}{4}( g_{ab}\nabla_c F_{ef} \nabla_d F^{ef} + g_{cd}\nabla_a F_{ef} \nabla_b F^{ef} )\nonumber\\
                & -\frac{1}{4} g_{ab}g_{cd}\nabla_e F_{fg} \nabla^e F^{fg}.
\end{align}
This tensor is completely symmetric in four dimensions, $H_{abcd} = H_{(abcd)}$ \cite{Bergqvist2003}. This tensor is
more interesting physically than the basic superenergy tensor, because it gives unique currents and a unique divergence
and because it shares the symmetries of the Bel tensor. We will now examine the currents that arise when this tensor is
contracted with the Killing vectors, $\xi^a$ and $\eta^a$, of our two-parameter group. Since the Chevreton tensor is
symmetric there are only four different currents, and by interchange of $\xi^a$ and $\eta^a$, we only need to consider
currents of the form
\begin{align} \label{ChevretonCurrentBeforeProof}
    H_{abcd}\xi^{Ib}\xi^{Jc}\xi^{Jd} =&
            -\nabla_a F_{ce}\nabla_b F_d{}^e \xi^{Ib}\xi^{Jc}\xi^{Jd} - \nabla_c F_{ae}\nabla_d F_b{}^e \xi^{Ib}\xi^{Jc}\xi^{Jd} \nonumber\\
           &+\frac{1}{2}\xi^J_c\xi^{Jc}\nabla_f F_{ae}\nabla^f F_b{}^e \xi^{Ib}
            +\frac{1}{4}\xi^J_c\xi^{Jc}\nabla_a F_{ef}\nabla_b F^{ef}\xi^{Ib} + \omega\xi_a + \Omega\eta_a,
\end{align}
Here and later $\omega$ and $\Omega$ are again used to collect the proportionality factors of $\xi_a$ and $\eta_a$. We
want to show that the remaining terms also lie in the orbits of the group. The proof is divided into three lemmas. We
treat the second and third terms separately and then the first and fourth together. The proofs involve some quite lengthy
calculations.

\begin{lemmap4} \label{lemma1}
    Under our assumptions,
    \begin{align}
        \xi_{[e}\eta_f \nabla^d F_{a]c}\nabla_d F_b{}^c\xi^{Ib} = 0.
    \end{align}
\end{lemmap4}
\begin{proof}
We rewrite $\nabla^d F_{ac}\nabla_d F_b{}^c\xi^{Ib}$ by applying the d'Alembertian $\Box = \nabla^d\nabla_d$ to the
energy-momentum tensor (\ref{EM-energymomentum})
\begin{align}
    \nabla^d F_{ac}\nabla_d F_b{}^c\xi^{Ib} = -\frac{1}{2}\xi^{Ib}( \Box T_{ab} + F_b{}^c\Box F_{ac} + F_a{}^c\Box F_{bc} ) + \omega\xi^I_a.
\end{align}
By using the four-dimensional Maxwell wave equation, $\Box F_{ab} = 2 C^c{}_{ab}{}^d F_{dc} - \frac{1}{3}R F_{ab}$
\cite{Penrose1986}, we get
\begin{align}
    -\frac{1}{2}\xi^{Ib}\Box T_{ab} - \xi^{Ib}( C^d{}_{ac}{}^e F_b{}^c F_{ed} + C^d{}_{bc}{}^e F_a{}^c F_{ed})
        +\frac{1}{3}R\xi^{Ib} F_{ac} F_b{}^c + \omega\xi^I_a.
\end{align}
By (\ref{Killing-energymomentum-prime}), the next to last term equals $\omega\xi_a + \Omega\eta_a$. The two terms
involving the Weyl tensor are rewritten using the four-dimensional identity \cite{Eriksson2006, Lovelock1970}
\begin{align}
    C_{[ab}{}^{[cd}\delta^{f]}_{e]} = 0.
\end{align}
We are then left with
\begin{align}
    -\frac{1}{2}\xi^{Ib}\Box T_{ab} - C_{ebad}F^{ec}F_c{}^d \xi^{Ib} + \omega\xi_a + \Omega\eta_a.
\end{align}
Substituting the Weyl tensor for Riemann tensor and simplifying with (\ref{Killing-energymomentum}) and Einstein's
equations (\ref{EinsteinEqs}) yields
\begin{align}
    -\frac{1}{2}\xi^{Ib}\Box T_{ab} - R_{ebad}F^{ec}F_c{}^d \xi^{Ib} + \omega\xi_a + \Omega\eta_a.
\end{align}
By (\ref{Killing-riemann}) we then have that
\begin{align} \label{temp-eq1}
    \nabla^d F_{ac}\nabla_d F_b{}^c\xi^{Ib} = -\frac{1}{2}\underbrace{\xi^{Ib}\Box T_{ab}}_{\mathcal{A}}
    - \underbrace{F^{ec}F_c{}^d \nabla_e\nabla_a\xi^I_d}_{\mathcal{B}} + \omega\xi_a + \Omega\eta_a.
\end{align}
We can rewrite the first term, $\mathcal{A}$, with the Leibniz rule as
\begin{align}
    \xi^{Ib}\Box T_{ab} = \Box(T_{ab}\xi^{Ib}) - T_{ab}\Box\xi^{Ib} -2 \nabla_f T_{ab}\nabla^f\xi^{Ib}.
\end{align}
For the first term on the right-hand side, use (\ref{Killing-energymomentum}), expand, and use
(\ref{Nablakilling-expanded}) and (\ref{Killing-scalars}). For the second term, use (\ref{Killing-riemann}) and
(\ref{Killing-energymomentum}). We are then left with
\begin{align}
    \xi^{Ib}\Box T_{ab} = -2\nabla_f T_{ab}\nabla^f\xi^{Ib} + \omega\xi_a +\Omega\eta_a.
\end{align}
Expanding this with (\ref{Nablakilling-expanded}) and rewriting with the Lie derivative of the energy-momentum tensor,
$\pounds_{\xi^I} T_{ab} =  \xi^{Ic}\nabla_c T_{ab} + T_{cb}\nabla_a\xi^{Ic} + T_{ac}\nabla_b\xi^{Ic} = 0$, and Leibniz
rule, we get
\begin{align}
    2x^{Ib}T_{fb}\nabla_a\xi^f + 2y^{Ib}T_{fb}\nabla_a\eta^f + 2x^{If}\nabla_f( T_{ab}\xi^b) + 2y^{If}\nabla_f( T_{ab}\eta^b)
        + \omega\xi_a + \Omega\eta_a.
\end{align}
Using (\ref{Killing-energymomentum}) and expanding again with (\ref{Nablakilling-expanded}) we are left with only
$\omega\xi_a + \Omega\eta_a$.

The second term of (\ref{temp-eq1}), $\mathcal{B}$, is expanded with (\ref{Nablakilling-expanded}) and using
(\ref{Killing-energymomentum-prime}) we get
\begin{align}
    F^{ec}F_c{}^d \nabla_e\nabla_a\xi^I_d = - (\alpha' \xi^e + \beta' \eta^e) \nabla_e x^I_a
                                            - (\gamma' \xi^e + \delta' \eta^e) \nabla_e y^I_a + \omega\xi_a + \Omega\eta_a.
\end{align}
Expanding $\pounds_{\xi^I} \nabla_a \xi^J_b=0$ gives $\pounds_{\xi^I}x^J_b = 0 = \pounds_{\xi^I}y^J_b$, or
$\xi^{Ia}\nabla_a x^J_b = \omega\xi_a + \Omega\eta_a$ and $\xi^{Ia}\nabla_a y^J_b = \omega\xi_a + \Omega\eta_a$. Hence,
\begin{align}
    F^{ec}F_c{}^d \nabla_e\nabla_a\xi^I_d = \omega\xi_a + \Omega\eta_a.
\end{align}
So, taken together,
\begin{align}
        \xi_{[e}\eta_f \nabla^d F_{a]c}\nabla_d F_b{}^c\xi^{Ib} = 0.
\end{align}

\end{proof}

As in our previous paper \cite{Eriksson2006}, we note here that lemma \ref{lemma1} can be applied to the trace of the
Chevreton tensor, which is given by \cite{Bergqvist2003}
\begin{align}
    H_{ab} = H_{abc}{}^c = \nabla_c F_{ad}\nabla^c F_b{}^d -\frac{1}{4}g_{ab}\nabla_c F_{de}\nabla^cF^{de}.
\end{align}
Hence
\begin{theoremp4}
    Assume that we have four-dimensional Einstein-Maxwell theory, possibly with a cosmological constant
    $\Lambda$, with a source-free electromagnetic field that inherits the symmetry of the spacetime. If $\xi_a$ and
    $\eta_a$ are two commuting Killing vectors that act orthogonally transitive on non-null surfaces, then the
    currents $H_{ab}\xi^b$ and $H_{ab}\eta^b$, where $H_{ab}$ is the trace of the Chevreton tensor, lie in the orbits
    of the group,
    \begin{align}
        H_{ab}\xi^b  &= \omega_1\xi_a + \Omega_1\eta_a, & H_{ab}\eta^b &= \omega_2\xi_a + \Omega_2\eta_a,
    \end{align}
    where the proportionality factors $\omega_i$ and $\Omega_i$ in general are non-constant.
\end{theoremp4}
These currents are trivially conserved, since the trace of the Chevreton tensor is divergence-free
\cite{Bergqvist2003}. Note that for a non-null electromagnetic field we automatically have inherited symmetry.

It was shown in \cite{Bergqvist2007} that the trace of the Chevreton tensor is related to the Bach tensor,
\begin{align}
    B_{ab} = \nabla^c \nabla^d C_{acbd} - \frac{1}{2}R^{cd}C_{acbd}.
\end{align}
by
\begin{align}
    B_{ab} = 2 H_{ab} + \frac{2}{3} \Lambda T_{ab}.
\end{align}
Hence, the Bach currents constructed from the Killing vectors $\xi_a$ and $\eta_a$ will also lie in the orbits of the
group. This also applies to the case with a hypersurface orthogonal Killing vector \cite{Eriksson2006}.
\begin{corollaryp4}
    Assume that we have four-dimensional Einstein-Maxwell theory, possibly with a cosmological constant
    $\Lambda$, with a source-free electromagnetic field that inherits the symmetry of the spacetime. If $\xi_a$ is a
    hypersurface orthogonal Killing vector, then the Bach current $B_{ab}\xi^b$ is proportional to $\xi_a$,
    \begin{align}
        B_{ab}\xi^b = \omega\xi_a.
    \end{align}
    If $\xi_a$ and $\eta_a$ are two commuting Killing vectors that act orthogonally transitive on non-null surfaces,
    then the Bach currents $B_{ab}\xi^b$ and $B_{ab}\eta^b$ lie in the orbits of the group
        \begin{align}
            B_{ab}\xi^b  &= \omega_3\xi_a + \Omega_3\eta_a,  &  B_{ab}\eta^b &= \omega_4\xi_a + \Omega_4\eta_a.
    \end{align}
    In general, the proportionality factors $\omega$, $\omega_i$, and $\Omega_i$ are non-constant.
\end{corollaryp4}
Again, in the second case, for a non-null electromagnetic field, we automatically have inherited symmetry for those
two Killing vectors generating the group.

For the proofs of the following two lemmas we will need to divide into two different cases depending on whether the
electromagnetic field is invertible or skew invertible. For a non-null electromagnetic field we can write
\cite{Michalski1975}
\begin{align}
    F_{ab} = \tau_{ab}\cos\alpha + \stackrel{*}\tau_{ab}\sin\alpha,
\end{align}
where $\tau_{ab}$ is the extremal field and $\alpha$ is the complexion scalar. The extremal field here satisfies one of
the following three sets of conditions \cite{Michalski1975}
\begin{align}
 &(1)   &   \tau_{ab}\xi^a\eta^b                &= 0       &       \stackrel{*}\tau_{ab}\xi^a\eta^b    &= 0, \\
 &(2)   &   \tau_{ab}\xi^a\eta^b                &\neq 0    &       \stackrel{*}\tau_{ab}\xi^a          &= 0 = \stackrel{*}\tau_{ab}\eta^a, \\
 &(3)   &   \stackrel{*}\tau_{ab}\xi^a\eta^b    &\neq 0    &       \tau_{ab}\xi^a                      &= 0 = \tau_{ab}\eta^a.
\end{align}
In the first case the electromagnetic field satisfies $F_{ab}\xi^a\eta^b =0= \stackrel{*}F_{ab}\xi^a\eta^b$, and is
said to be skew invertible. It can then be written as
\begin{align} \label{Em-skewinvertible}
    F_{ab} =2\xi_{[a}s_{b]} + 2\eta_{[a}t_{b]},
\end{align}
where $s_a$ and $t_a$ are orthogonal to $\xi_a$ and $\eta_a$. Carter \cite{Carter1969} showed that the two scalars
$F_{ab}\xi^a\eta^b$ and $\stackrel{*}F_{ab}\xi^a\eta^b$ are constants and if we, for example, have a spacetime with a
symmetry axis where one of the Killing vectors vanishes, the constants vanish everywhere and the electromagnetic field
will be skew invertible. In the two other cases the electromagnetic field is invertible and can be written as
\begin{align}
    F_{ab} = 2\kappa \xi_{[a}\eta_{b]} + 2s_{[a}t_{b]},
\end{align}
where again $s_a$ and $t_a$ are orthogonal to $\xi_a$ and $\eta_a$.

For a null electromagnetic field with principal null direction $l_a$ we can write
\begin{align}
    F_{ab} &= 2l_{[a}A_{b]},    &   \stackrel{*}F_{ab} = 2l_{[a}B_{b]},
\end{align}
where $A_a$ and $B_a$ are spacelike vectors satisfying $A^a l_a = B^a l_a = A^a B_a = 0$. By expanding
(\ref{Killing-energymomentum-prime}) we see that either $l_a = \omega\xi_a+\Omega\eta_a$ or $\xi^a l_a =0= \eta^a l_a$,
which in either case implies $F_{ab}\xi^a\eta^b = 0 = \stackrel{*}F_{ab}\xi^a\eta^b$ and the electromagnetic field is
therefore skew invertible.

In the following two proofs we will only show the calculations for the skew invertible case. The invertible case works
similarly, noting that $F_{ab}\xi^{Ib} = \lambda^I\xi_a + \mu^I\eta_a$, where $\xi^{Ia}\nabla_a\lambda^J = 0$ and
$\xi^{Ia}\nabla_a\mu^J = 0$.

\begin{lemmap4} \label{lemma2}
    Under our assumptions,
    \begin{align}
        \xi_{[f}\eta_g \nabla_{|c|} F_{a]e} \nabla_d F_b{}^e \xi^{Ib}\xi^{Jc}\xi^{Jd} = 0.
    \end{align}
\end{lemmap4}

\begin{proof}
We start by rewriting with the Lie derivative (\ref{EM-Liederivative}),
\begin{align} \label{temp-eq2}
    \nabla_c F_{ae} \nabla_d F_b{}^e \xi^{Ib}\xi^{Jc}\xi^{Jd} =&
          \xi^{Ib}\nabla_a\xi^{Jc}\left( F_{ce}F_d{}^e \nabla_b \xi^{Jd} + F_{ce}F_{bd}\nabla^e \xi^{Jd}\right) \nonumber\\
        &+\xi^{Ib}F_{ac}\left( F_d{}^e \nabla_e\xi^{Jc}\nabla_b\xi^{Jd} + F_{bd}\nabla_e\xi^{Jc}\nabla^e\xi^{Jd}\right).
\end{align}
Expanding the first term of the right-hand side with (\ref{Nablakilling-expanded}) and using
(\ref{Killing-energymomentum-prime}), we have
\begin{align}
    \xi^{Ib}\nabla_a\xi^{Jc} F_{ce}F_d{}^e \nabla_b \xi^{Jd} = \omega\xi_a + \Omega\eta_a.
\end{align}
The second term of (\ref{temp-eq2}) is expanded by (\ref{Nablakilling-expanded}), and if the electromagnetic field is
skew invertible, all terms like $F_{ab}\xi^{Ka}\xi^{Lb}$ vanish. Hence,
\begin{align}
    \xi^{Ib}\nabla_a\xi^{Jc}F_{ce}F_{bd}\nabla^e \xi^{Jd} =\omega\xi_a + \Omega\eta_a.
\end{align}
For the third term, using (\ref{Em-skewinvertible}) and (\ref{KillingKillingNablaKilling}) we have
\begin{align}
    \xi^{Ib}F_{ac} F_d{}^e \nabla_e\xi^{Jc}\nabla_b\xi^{Jd} &= -s_a \xi_c \xi^{Ib} F_d{}^e\nabla_e\xi^{Jc}\nabla_b\xi^{Jd}
                                                              -t_a \eta_c \xi^{Ib} F_d{}^e\nabla_e\xi^{Jc}\nabla_b\xi^{Jd}
                                                              +\omega\xi_a + \Omega\eta_a \nonumber\\
                                                            &=\omega\xi_a + \Omega\eta_a.
\end{align}
For the last term of (\ref{temp-eq2}) we expand using (\ref{Nablakilling-expanded}) and (\ref{Em-skewinvertible}) to get
\begin{align}
    \xi^{Ib}F_{ac} F_{bd}\nabla_e\xi^{Jc}\nabla^e\xi^{Jd} = \omega\xi_a + \Omega\eta_a.
\end{align}
Hence, taken together, we have that
\begin{align}
        \xi_{[f}\eta_g \nabla_{|c|} F_{a]e} \nabla_d F_b{}^e \xi^{Ib}\xi^{Jc}\xi^{Jd} = 0.
\end{align}
The proof is similar for the invertible electromagnetic field.

\end{proof}

\begin{lemmap4} \label{lemma3}
    Under our assumptions,
    \begin{align}
            -\xi_{[g}\eta_h \nabla_{a]} F_{ce}\nabla_b F_d{}^e \xi^{Ib}\xi^{Jc}\xi^{Jd}
            +\frac{1}{4}\xi^J_c\xi^{Jc}  \xi_{[g}\eta_h \nabla_{a]} F_{ef}\nabla_b F^{ef}\xi^{Ib} =0.
    \end{align}
\end{lemmap4}

\begin{proof}
Taking two covariant derivatives of the energy-momentum tensor (\ref{EM-energymomentum}) yields
\begin{align}
    -\nabla_a F_{ce}\nabla_b F_d{}^e \xi^{Ib}\xi^{Jc}\xi^{Jd} + \frac{1}{4}\xi^J_c\xi^{Jc}  \nabla_a F_{ef}\nabla_b F^{ef}\xi^{Ib} = \nonumber\\
        \frac{1}{2}\underbrace{\nabla_a\nabla_b T_{cd}\xi^{Ib}\xi^{Jc}\xi^{Jd}}_\mathcal{A}
        + \underbrace{F_{ce}\nabla_a\nabla_b F_d{}^e\xi^{Ib}\xi^{Jc}\xi^{Jd}}_\mathcal{B}
        -\frac{1}{4}\xi^{Jc}\xi^J_c\underbrace{\xi^{Ib}F_{ef}\nabla_a\nabla_b F^{ef}}_\mathcal{C}.
\end{align}
We rewrite term $\mathcal{A}$ with a covariant derivative of the Lie derivative of the energy-momentum tensor,
$\nabla_a \pounds_\xi T_{cd} = 0$,
\begin{align}
    \nabla_a\nabla_b T_{cd}\xi^{Ib}\xi^{Jc}\xi^{Jd} =
        -\nabla_a\xi^{Ib}\nabla_b T_{cd}\xi^{Jc}\xi^{Jd}
        -2\nabla_a T_{bc} \nabla_d \xi^{Ib} \xi^{Jc}\xi^{Jd}
        -2 T_{bc}\nabla_a\nabla_d \xi^{Ib} \xi^{Jc}\xi^{Jd}.
\end{align}
By expanding with (\ref{Nablakilling-expanded}), the Lie derivative of energy-momentum tensor,
(\ref{Killing-energymomentum}), and (\ref{KillingKillingNablaKilling}),
\begin{align}
    \nabla_a\xi^{Ib}\nabla_b T_{cd}\xi^{Jc}\xi^{Jd} = \omega\xi_a + \Omega\eta_a.
\end{align}
Similarly, by using Leibniz rule, expanding with (\ref{Nablakilling-expanded}) and using (\ref{Killing-energymomentum})
and (\ref{KillingKillingNablaKilling})
\begin{align}
    \nabla_a T_{bc} \nabla_d \xi^{Ib} \xi^{Jc}\xi^{Jd} = \omega\xi_a + \Omega\eta_a.
\end{align}
Finally, by expanding with (\ref{Nablakilling-expanded}) twice and using (\ref{Killing-energymomentum}), Leibniz rule,
and (\ref{KillingKillingNablaKilling}),
\begin{align}
    T_{bc}\nabla_a\nabla_d \xi^{Ib} \xi^{Jc}\xi^{Jd} = \omega\xi_a + \Omega\eta_a.
\end{align}
Term $\mathcal{B}$, rewritten by taking a covariant derivative of the Lie derivative of the electromagnetic field
(\ref{EM-Liederivative}) equals
\begin{align}
    F_{ce}\nabla_a\nabla_b F_d{}^e\xi^{Ib}\xi^{Jc}\xi^{Jd} =&
        -F_c{}^e\nabla_a F_{be}\nabla_d \xi^{Ib} \xi^{Jc}\xi^{Jd}
        -F_c{}^e\nabla_a F_{db}\nabla_e \xi^{Ib} \xi^{Jc}\xi^{Jd}\nonumber\\
      &  -F_c{}^e\nabla_a\xi^{Ib}\nabla_b F_{de} \xi^{Jc}\xi^{Jd}
        -F_c{}^eF_{be}\nabla_a\nabla_d \xi^{Ib} \xi^{Jc}\xi^{Jd}
\end{align}
The first term is expanded with (\ref{Nablakilling-expanded}) and if the electromagnetic field is skew invertible, we
use (\ref{Em-skewinvertible}), to yield
\begin{align}
        -F_c{}^e\nabla_a F_{be}\nabla_d \xi^{Ib} \xi^{Jc}\xi^{Jd} = \omega\xi_a + \Omega\eta_a.
\end{align}
The second term is similarly expanded with (\ref{Nablakilling-expanded}) and (\ref{Em-skewinvertible}) and then Leibniz
rule is used to give us
\begin{align}
        -F_c{}^e\nabla_a F_{db}\nabla_e \xi^{Ib} \xi^{Jc}\xi^{Jd} = \omega\xi_a + \Omega\eta_a.
\end{align}
The third term is expanded with (\ref{Nablakilling-expanded}) to yield
\begin{align}
    - F_c{}^e\nabla_a\xi^{Ib}\nabla_b F_{de} \xi^{Jc}\xi^{Jd} =&  x^I_a F_c{}^e \xi^b \nabla_b F_{de}\xi^{Jc}\xi^{Jd}
                                                                 +y^I_a F_c{}^e \eta^b \nabla_b F_{de}\xi^{Jc}\xi^{Jd} +\omega\xi_a + \Omega\eta_a \nonumber\\
                                                              =&\frac{1}{2} x^I_a  \xi^b \nabla_b(F_c{}^e F_{de})\xi^{Jc}\xi^{Jd}
                                                               +\frac{1}{2} y^I_a  \eta^b \nabla_b(F_c{}^e F_{de})\xi^{Jc}\xi^{Jd}+\omega\xi_a + \Omega\eta_a,
\end{align}
where, by (\ref{Killing-energymomentum-prime}), (\ref{Killing-scalars}), and (\ref{KillingKillingNablaKilling}) we have
\begin{align}
    \xi^b \nabla_b(F_c{}^e F_{de})\xi^{Jc}\xi^{Jd} =& \xi^b \nabla_b(F_c{}^e F_{de}\xi^{Jc})\xi^{Jd}- F_c{}^e F_{de}\xi^b \nabla_b\xi^{Jc}\xi^{Jd} \nonumber\\
                =& \xi^b\xi^{Jd} \nabla_b(\alpha'^J\xi_d + \beta'^J\eta_d) - (\alpha'^J\xi_d + \beta'^J\eta_d)\xi^b \nabla_b\xi^{Jd} =0,
\end{align}
and likewise for the other term. Hence,
\begin{align}
    - F_c{}^e\nabla_a\xi^{Ib}\nabla_b F_{de} \xi^{Jc}\xi^{Jd} = \omega\xi_a + \Omega\eta_a.
\end{align}
For the last term we use (\ref{Killing-energymomentum-prime}), expand with (\ref{Nablakilling-expanded}), and Leibniz
rule to get
\begin{align}
    - F_c{}^eF_{be}\nabla_a\nabla_d \xi^{Ib} \xi^{Jc}\xi^{Jd} = \omega\xi_a + \Omega\eta_a.
\end{align}
Term $\mathcal{C}$, rewritten by taking a covariant derivative of Lie derivative of the electromagnetic field, equals
\begin{align}
    F_{ef}\nabla_a\nabla_b F^{ef}\xi^{Ib} =&
        2F^{ef}\nabla_a F_{be}\nabla_f \xi^{Ib} - F^{ef}\nabla_a\xi^{Ib}\nabla_b F_{ef}
\end{align}
The first term is expanded with (\ref{Nablakilling-expanded}) and (\ref{Em-skewinvertible}) to yield
\begin{align}
        F^{ef}\nabla_a F_{be}\nabla_f \xi^{Ib} = \omega\xi_a+\Omega\eta_a
\end{align}
The second term is expanded with (\ref{Nablakilling-expanded}) and rewritten with the Lie derivative
(\ref{EM-Liederivative}),
\begin{align}
    F^{ef}\nabla_a\xi^{Ib}\nabla_b F_{ef} = -x^I_a F^{ef}\xi^b\nabla_b F_{ef} - y^I_a F^{ef}\eta^b\nabla_b F_{ef} + \omega\xi_a+\Omega\eta_a = \omega\xi_a+\Omega\eta_a
\end{align}
Hence, taken together, we have that
\begin{align}
            -\xi_{[g}\eta_h \nabla_{a]} F_{ce}\nabla_b F_d{}^e \xi^{Ib}\xi^{Jc}\xi^{Jd}
            +\frac{1}{4}\xi^J_c\xi^{Jc}  \xi_{[g}\eta_h \nabla_{a]} F_{ef}\nabla_b F^{ef}\xi^{Ib} =0.
\end{align}
The case with an invertible electromagnetic field works similarly.
\end{proof}

From (\ref{ChevretonCurrentBeforeProof}) together with lemmas \ref{lemma1}, \ref{lemma2}, and \ref{lemma3} we
have that $H_{abcd}\xi^{Ib}\xi^{Jc}\xi^{Kd} = \omega_{IJK}\xi_a + \Omega_{IJK}\eta_a$. The Lie derivative commutes with
the covariant derivative for Killing vectors (\ref{LieCovariant}), so $\pounds_{\xi^I} \nabla_a F_{bc} = 0$ and we have
that $\pounds_{\xi^I} H_{abcd} = 0$. Since the Killing vectors commute, we have that
\begin{align}
    \pounds_{\xi^I} (H_{abcd} \xi^{Jb}\xi^{Kc}\xi^{Ld}) = 0.
\end{align}
Hence, the proportionality factors $\omega$ and $\Omega$ satisfies
$\xi^{Ia}\nabla_a\omega = 0 = \xi^{Ia}\nabla_a\Omega$. We have proven

\begin{theoremp4} \label{BigTheorem}
    Assume that we have four-dimensional Einstein-Maxwell theory, possibly with a cosmological constant, $\Lambda$,
    with a source-free electromagnetic field that inherits the symmetry of the spacetime.
    If $\xi^a$ and $\eta^a$ generate a two-parameter Abelian isometry group that act orthogonally transitive on
    non-null surfaces, then the Chevreton currents constructed from these vectors lie in the orbits of the group and
    are divergence-free,
    \begin{align}
        H_{abcd}\xi^b\xi^c\xi^d    &= \omega_5\xi_a+\Omega_5\eta_a,  &   \nabla^a(H_{abcd}\xi^b\xi^c\xi^d)    &=0. \nonumber\\
        H_{abcd}\xi^b\xi^c\eta^d   &= \omega_6\xi_a+\Omega_6\eta_a,  &   \nabla^a(H_{abcd}\xi^b\xi^c\eta^d)   &=0.\nonumber\\
        H_{abcd}\xi^b\eta^c\eta^d  &= \omega_7\xi_a+\Omega_7\eta_a,  &   \nabla^a(H_{abcd}\xi^b\eta^c\eta^d)  &=0.\nonumber\\
        H_{abcd}\eta^b\eta^c\eta^d &= \omega_8\xi_a+\Omega_8\eta_a,  &   \nabla^a(H_{abcd}\eta^b\eta^c\eta^d) &=0.
    \end{align}
    In general, the proportionality factors $\omega_i$ and $\Omega_i$ will be non-constant.
\end{theoremp4}
Note that for a non-null electromagnetic field, the symmetry is automatically inherited.

In four-dimensional Einstein-Maxwell theory, the Bel tensor (\ref{BelTensor}) can be decomposed as
\cite{Bonilla1997, Eriksson2006}
\begin{align}
    B_{abcd} = T_{abcd} + T_{ab}T_{cd} + \frac{1}{48}R^2 g_{ab}g_{cd},
\end{align}
where $T_{abcd}$ is the Bel-Robinson tensor (\ref{BRTensor}) and $T_{ab}$ is the electromagnetic energy-momentum
tensor (\ref{EM-energymomentum}). As was shown in \cite{Lazkoz2003}, the Bel currents
$B_{a(bcd)}\xi^{Ib}\xi^{Jc}\xi^{Kd}$ lie in the orbits of the group. We note that when we contract with the Killing
vectors, both the second and the third terms above will lie in the orbits of the group. Both terms also have vanishing
Lie derivative, so both terms will give rise to independently conserved currents. From this we see that this also
applies to the Bel-Robinson tensor, i.e., it will also give rise to conserved currents that lie in the orbits of the
group.


\section{Example}
Theorem \ref{BigTheorem} applies to axisymmetric spacetimes, notably the Kerr-Newman solution. The expressions for the
currents are however quite large, so we give instead another interesting example -- the algebraically general
Einstein-Maxwell spacetime found by Barnes \cite{Barnes1976}. The metric is given by
\begin{align}
    {\rm d}s^2 = r\sin(\sqrt{3}\theta){\rm d}x^2 - r\sin(\sqrt{3}\theta){\rm d}y^2 - r2\cos(\sqrt{3}\theta){\rm d}x{\rm d}y
                    -r^2{\rm d}\theta^2 - {\rm d}r^2,
\end{align}
and it admits a three-parameter group of isometries generated by the Killing vectors
\begin{align}
    \xi_{1a} &= \delta_{xa},     &       \xi_{2a} &= \delta_{ya},   &       \xi_{3a} &=
            y\frac{\sqrt{3}}{2}\delta_{xa} - x\frac{\sqrt{3}}{2}\delta_{ya} + \delta_{\theta a}.
\end{align}
Here $\xi_1$ and $\xi_2$ form an orthogonally transitive Abelian $G_2$ subgroup. None of the three Killing vectors are
hypersurface orthogonal. The electromagnetic field is given by
\begin{align}
    F_{ab} = 2\cos(p) \delta_{xa}\wedge\delta_{yb} + 2\sin(p) \delta_{ra}\wedge\delta_{\theta b},
\end{align}
where $p$ is an arbitrary constant determining the complexion of the field.

The four Chevreton and Bel currents constructed from $\xi_1$ and $\xi_2$ are of course
independently conserved. The Chevreton currents are here very simple,
\begin{align}
    H_{abcd} \xi_1^b\xi_1^c\xi_1^d &= \frac{3}{2r^2}\xi_{1a},       &      H_{abcd} \xi_1^b\xi_1^c\xi_2^d &= \frac{1}{2r^2}\xi_{2a},      \nonumber\\
    H_{abcd} \xi_1^b\xi_2^c\xi_2^d &= \frac{1}{2r^2}\xi_{1a},       &      H_{abcd} \xi_2^b\xi_2^c\xi_2^d &= \frac{3}{2r^2}\xi_{2a}.
\end{align}
None of the six Chevreton currents involving the third Killing vector $\xi_3$ are divergence-free and the same holds
for the corresponding Bel currents. However, a combination of the symmetrized Bel tensor with the Chevreton tensor do
give conserved currents for all possible combinations of the Killing vectors,
\begin{align}
    \nabla^a \left( (B_{a(bcd)}+ \frac{1}{3}H_{abcd})\xi^{b}_i\xi^{c}_j\xi^{d}_k\right) = 0.
\end{align}
Actually, for this spacetime, all Bel-Robinson currents are independently conserved and it is only the $T_{ab}T_{cd}$
part of the Bel tensor that contributes to the mixed current. For example, the current
$v_a = (B_{a(bcd)}+ \frac{1}{3}H_{abcd})\xi^{b}_1\xi^{c}_2\xi^{d}_3$, which equals
\begin{align}
   v_x &= -\frac{\sqrt{3}\left(3x\cos(2\sqrt{3}\theta)-3y\sin(2\sqrt{3}\theta) +11x\right)}{48r^2}, \nonumber\\
   v_y &= \frac{\sqrt{3}\left(3y\cos(2\sqrt{3}\theta)+3x\sin(2\sqrt{3}\theta) +11y\right)}{48r^2}, \nonumber\\
   v_r &= \frac{\sqrt{3}\sin(\sqrt{3}\theta)}{6r^2}, \nonumber\\
   v_\theta &= \frac{\cos(\sqrt{3}\theta)}{3r},
\end{align}
is conserved. It is also interesting to note that this current is not proportional to a combination of the Killing
vectors. This actually holds for all the currents here that involve the third Killing vector $\xi_3$.


\section{Conclusion}

We have shown that if a four-dimensional Einstein-Maxwell spacetime admits an Abelian two-parameter isometry group
that act orthogonally transitive on non-null surfaces and the electromagnetic field is source-free and inherits the
symmetries of the spacetime, then the Chevreton currents generated from the isometry group lies in the orbits of the
group and are conserved. Hence, by Gauss's theorem these currents give rise to conserved quantities.

Since the Bel currents have similar properties under this isometry group, this gives further support to the possibility
of constructing mixed conserved currents that could govern the interchange of superenergy between the electromagnetic
field and the gravitational field.

In the proof of lemma \ref{lemma1} we needed to make use of an identity which holds only in four dimensions, so our
result seems to be restricted to this dimension. The results for the Bel currents are $n$-dimensional, so possible
mixed conservation laws may be restricted to four dimensions.


\section*{Acknowledgements}
The author wishes to thank Göran Bergqvist and José Senovilla for valuable comments and discussions.


\end{document}